\documentclass{IEEEtran}
\usepackage{amssymb}
\usepackage{amsfonts}
\usepackage{amsmath}
\usepackage{algorithm}
\usepackage{graphicx,subfigure,cite,multicol,multirow,diagbox,booktabs,array}

\renewcommand{\arraystretch}{1.5}
\usepackage[dvips]{color}
\usepackage{float}
\ifCLASSINFOpdf
\else
\fi
\begin{document}
\title{Fast Ambiguous DOA Elimination Method of DOA Measurement for Hybrid Massive MIMO Receiver}

\author{Nuo Chen, Xinyi Jiang, Baihua Shi, Yin Teng, Jinhui Lu, Feng Shu, Jun Zou, Jun Li, and Jiangzhou Wang

}
\maketitle

\begin{abstract}
DOA estimation for massive multiple-input multiple-output (MIMO) system can provide ultra-high-resolution angle estimation. However, due to the high computational complexity and cost of all digital MIMO systems, a hybrid analog digital (HAD) structure MIMO was proposed. In this paper, a fast ambiguous phase elimination method is proposed to solve the problem of direction-finding ambiguity caused by the HAD MIMO. Only two-data-blocks are used to realize DOA estimation. Simulation results show that the proposed method can greatly reduce the estimation delay with a slight performance loss.
\end{abstract}


\IEEEpeerreviewmaketitle

\section{Introduction}
Direction of Arrival (DOA) estimation has been widely used in many applications, including wireless communications, radar, navigation, sonar, tracking of various objects, secure and precise wireless transmission
(SPWT), rescue and other emergency assistance equipment \cite{2009Tuncer}, \cite{2018Shu}. In recent years, DOA estimation for massive multiple-input multiple-output (MIMO) system has attracted a lot of attention, which can provide ultra-high-resolution angle estimation. A novel framework of combining deep-learning and massive MIMO  was proposed in \cite{2018Huang} to realize super-resolution channel estimation and DOA estimation. However, as the number of antennas tends to large-scale, due to its high computational complexity and circuit cost, it is difficult for massive MIMO to be widely used in DOA measurement.  To address this issue, in \cite{2018shuqin}, a hybrid analog digital (HAD) structure MIMO was proposed, multiple low-complexity phase alignment(PA) methods were proposed to estimate DOA, and the corresponding Cramer-Rao lower bound(CRLB) is derived. A novel DOA-aided channel estimation for a HAD MIMO precoding system at the base station (BS) was proposed in \cite{2018Fan}  to  achieve the CRLB. A low-complexity deep-learning-based DOA estimation method for a HAD MIMO system  was proposed in \cite{2019Hu}, which can achieve similar or even lower the normalized mean square error (NMSE) with much less complexity compared to the maximum likelihood (ML) method. For a HAD MIMO structure, the DOA measurement process falls into two stages: DOA estimation of generating a set of candidate solutions and  cancelling spurious solutions. For a HAD MIMO, the major challenging problem is how to eliminate direction-finding ambiguity rapidly. A smart strategy of maximizing the average receive power was proposed to remove $M-1$ spurious solutions in \cite{2018shuqin}, where $M$ is the number of antennas per subarray. This means it requires about $M-1$ time slots to infer the true direction angle with each time slot being multiple snapshots or samples. This means a large processing delay of $M$ time slots. In this paper, a fast ambiguous phase elimination method is proposed to find the true solution using only two-data-blocks by exploring the HAD structure with a slight performance  loss.

\section{System Model}

The HAD antenna array captures the narrowband signal $s(t)e^{j2\pi f_ct}$ from the $\theta_0 $ direction emitted by a far-field transmitter, where $s(t)$ is the baseband signal and $f_c$ is the carrier frequency. Here, a uniformly-spaced linear array (ULA) with $N$ antennas is deployed and divided into $K$ subarrays with each subarray containing $M$ antennas where $N=MK$. Via analog beamforming (AB), radio frequency (RF) chains, analog-to-digital convertor (ADC) and digital beamforming (DB), the resulting receive signal is  $r^b(n)=\mathbf{v}^H_D\mathbf{V}^H_A\mathbf{a}(\theta_0)s(n)+\mathbf{v}^H_D\mathbf{V}^H_A\mathbf{w}^b(n)$, where $b$ denotes the index of time slots, each time slot consists of $L$ snapshots, $\mathbf{w}^b(n)\sim \mathcal{C}\mathcal{N}(0,\sigma^2_w\mathbf{I}_M)$ is an additive white Gaussian noise (AWGN), $\mathbf{a}(\theta_0)$ is an array manifold, the DB vector is $\mathbf{v}_D=[v_1,v_2,\cdots,v_K]^T$, and the AB matrix $\mathbf{V}_{A}$ is a block diagonal matrix. Let us define $\varphi=\frac{2\pi}{\lambda}d\sin{\theta_0}$, where $\lambda$ represents the signal wavelength and $d$ represents the antenna spacing.
\section{Conventional Root-MUSIC-HDAPA DOA Estimator}
In the first stage, when all AB phases  are zero, the output vector of sample $n$ in time slot $b$ is $ \mathbf{y}^{b}_{AB}(n)=M^{-\frac{1}{2}}\mathbf{a}_D(\theta_0)s^b(n)+\mathbf{w}^b_{AB}(n)$, where $\mathbf{a}_D(\theta_0)=g(\theta_0)\mathbf{a}_M(\theta_0)$, $\mathbf{a}_M(\theta_0)=[1,e^{jM\varphi},\cdots,e^{j(K-1)M\varphi}]^T$,
$ g(\theta_0)=\sum\limits_{m=1}^{M}e^{j(m-1)\varphi}$.
The set of candidate solutions to DOA is estimated by using the Root-MUSIC algorithm. The sample covariance matrix of the output vector of the antenna array is $\mathbf{R}^b_{yy}=1/L\sum^{L}_{n=1}\mathbf{y}^b_{AB}(n)\mathbf{y}^b_{AB}(n)
$, whose singular-value decomposition (SVD)
is expressed as $\mathbf{R}_{yy}=[\mathbf{E}_S\ \mathbf{E}_N]\sum[\mathbf{E}_S\ \mathbf{E}_N]^H$ where $\mathbf{E}_S$ and $\mathbf{E}_N$ correspond to signal and noise subspaces, respectively.
so the corresponding spectral function is $ P_{MU}(\theta)=\|\mathbf{a}^H_D(\theta)\mathbf{E}_N\mathbf{E}^H_N\mathbf{a}_D(\theta)\|^{-1}$. Let us define the polynomial equation: $f_\theta (\theta)=\mathbf{a}^H_D(\theta)\mathbf{E}_N\mathbf{E}^H_N\mathbf{a}_D(\theta)\triangleq f_z(z)\triangleq f_{\phi}(\phi)=0$, where $z=e^{jM\varphi} $, and $\phi=M\varphi$. The polynomial equation $f_z(z)$ has  $2K-2$ roots $z_i$, which yields a set of associated emitter phases $\hat{\Theta}_r=\{\hat{\phi}_{r,i},i\in\{1,2,\cdots,2K-2\}\}$. Digital phase alignment (DPA) is used to delete $2K-3$ pseudo solutions in $\hat{\Theta}_r$ and $\hat{\phi}_r$ is obtained. Then we can get $\hat{\phi}_r=2\pi\lambda^{-1}Md\sin{\hat{\theta}_r}$. Since the function $f_\phi(\phi)$ is a periodic function of $\phi$  with period $2\pi$, therefore, the extended feasible solution set is $\hat{\Theta}=\{\hat{\theta}_i,i\in \{0,1,\cdots,M-1\}\}$, where $ \hat{\theta}_i=\arcsin(\frac{\lambda(\hat{\phi}_r+2\pi i)}{2\pi Md})$. Finally, analog phase alignment (APA) is used to eliminate the spurious solutions in the feasible set $\hat{\Theta}$ .

Considering the analog signal cannot be stored before ADC, the new $M-1$ time slots should be received to eliminate $M-1$ spurious direction ambiguity  in \cite{2018shuqin}. This will lead to a large estimation delay. To address this problem, a fast ambiguous phase elimination method is proposed to eliminate the spurious solutions by using only single time slot.

\section{Proposed fast method of removing spurious solutions}

\begin{figure}[H]
\centering
\includegraphics[width=0.42\textwidth]{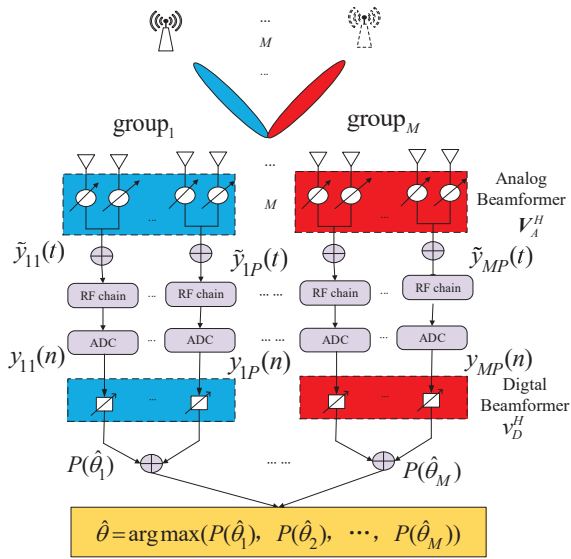}
\caption{Proposed fast structure of removing spurious direction angles.}
\label{RMSE_SNR}
\end{figure}
 Fig.~1 shows the basic idea of eliminating spurious directions with $K\geqslant M$ in the second time slot. The total number $K$ of subarrays are categorized into $M$ groups of subarrays where each group has $P=K/M$ subarrays.  In this slot, the phases of receive APA  are designed according to $M$ ambiguous directions such that all phases of the subarray group corresponding to the true direction are aligned to output the maximum power after APA, the output  signal of the $p$th subarray of  group $m$ is as follows:
\begin{align}
y_{mp}(n)=\mathbf{v}^H_{A,mp}\mathbf{a}_{mp}(\theta_0)s(n)+w_{mp}(n)
\end{align}
where
\begin{align}
\mathbf{v}_{A,mp}=\frac{1}{\sqrt{M}}[e^{j\alpha_{mp,0}},e^{j\alpha_{mp,1}},\cdots,e^{j\alpha_{mp,M-1}}]
\end{align}
where
\begin{align}
\alpha_{mp,i}=\frac{2\pi}{\lambda}(H+i)d\sin\hat{\theta}_m
\end{align}
where $H=(m-1)PM+(P-1)M$. The DB vector is set to be $\mathbf{v}_D=[1,1,\cdots,1]^T$. Therefore, the output signal through DPA is $ r_m(n)=\sum\limits_{p=1}^{P}y_{mi}(n)$, and the average output power is
\begin{align}
P_r(\hat{\theta}_m)=\frac{1}{L}\sum\limits_{n=1}^{L}[r_m(n)r_m(n)^H]=\frac{1}{L}\mathbf{r}\mathbf{r}^H
\end{align}
where $\mathbf{r}=[r(1),\cdots,r(L)]$. Eventually, the true direction angle corresponding to the maximum average power is
\begin{align}
\hat{\theta}=\mathop{\arg\max_{\hat{\theta}_m\in\hat{\Theta}}}\ P_r(\hat{\theta}_m)
\end{align}
which completes the cancellation of spurious angles of requiring only one time slot. The delay is significantly  reduced compared to the existing Root-MUSIC-HDAPA DOA Estimator method in  \cite{2018shuqin}. The ratio of their total time delays  is $2/(M+1)$. As $M$ increases, the rapid advantage of  the proposed method over Root-MUSIC-HDAPA is more dramatic.

\section{Computational complexity analysis}
The computational complexities of the existing method and proposed method are $C_{original}=\mathcal{O}(K^2L+(2(K-1))^3+L((2K-2)K+NM))$, $C_{proposed}=\mathcal{O}(K^2L+(2(K-1))^3+L((2K-2)K+N))$ float-point operations (FLOPs). We can know that the computational complexity of the proposed method is reduced by $M$ times when the ambiguous phase is eliminated.

\section{Simulation and Discussion}
System parameters are chosen as follows: the direction of emitter $\theta_0=41.345^{\circ}$, $N=64$, $M=4$, $L=8$.

\begin{table}[H]
\renewcommand{\arraystretch}{1.3}
\caption{RMSE versus SNR with $N=64$}
\label{tab1}
\centering
\resizebox{0.47\textwidth}{!}{
\begin{tabular}{p{4cm}|c|c|c|c|c}
\hline SNR(dB)&-20&-15&-10&-5&0\\
\hline RMSE(Proposed method)&32.9&29.9&21.5&7.2&0.19\\
\hline RMSE(Root-MUSIC-HDAPA)&29.3&25.3&18.4&5.7&0.18\\
\hline
\end{tabular}}
\end{table}

Table~\ref{tab1} illustrates the  performances of root mean square error (RMSE) versus SNR  of  the proposed method and existing Root-MUSIC-HDAPA in \cite{2018shuqin}. It can be seen from Table~\ref{tab1} that the proposed method is slightly worse than Root-MUSIC-HDAPA in terms of RMSE due to the use of far much less samples.

\ifCLASSOPTIONcaptionsoff
  \newpage
\fi

\bibliographystyle{IEEEtran}
\bibliography{IEEEfull,cite}

\end{document}